\documentclass[a4paper,11pt]{article}
\usepackage{AASKAII_draft_template_v1/aaskaiid}
\newcommand{\apjl}{ApJL} 
\newcommand{\apjs}{ApJSS} 
\newcommand{\ao}{Appl. Opt.} 
\newcommand{\aap}{A\&A} 

\graphicspath{{AASKAII_draft_template_v1/}{./}}
\usepackage{caption}
\usepackage{xcolor}

\usepackage{orcidlink}
\usepackage{tabularx}   
\usepackage{array}      
\usepackage{booktabs}   
 \usepackage{siunitx}
\newcolumntype{Y}{>{\raggedleft\arraybackslash}X} 

\usepackage{aas_macros} 

\newcommand{\jw}[1]{\textcolor{black}{#1}}
\newcommand{\mk}[1]{\textcolor{orange}{Matthias: #1}}

\title{VLBI with SKAMPI, the SKA-Mid MPIfR dish demonstrator}
\ShortTitle{SKAMPI VLBI}

\author[1,2]{Jompoj Wongphechauxsorn
\orcidlink{0000-0002-7730-4956}}

\ShortName{Wongphechauxsorn et al.} 
\author[2]{Niclas Alexander Esser}
\author[2]{Tobias Winchen}
\author[2]{Jan Wagner\orcidlink{0000-0003-1105-6109}}
\author[2]{Uwe Bach}
\author[2]{Hans-Rainer Kl\"ockner\orcidlink{0000-0002-0648-2704}}
\author[2]{Michael Kramer}
\author[2]{Gundolf Wieching}
\author[2]{Ewan Barr}
\author[3]{Jonathan Quick}
\author[4]{Roger Deane}
\author[5]{Cormac Reynolds}
\author[5]{Phil Edwards}
\author[5]{Chris Phillips\orcidlink{0000-0002-5851-5264}}
\author[1]{Matthias Kadler \orcidlink{0000-0001-5606-6154}}
\author[6]{Roopesh Ojha}

\emailAdd{jompoj.wongphechauxsorn@uni-wuerzburg.de}
\emailAdd{hrk@mpifr-bonn.mpg.de}

\affiliation[1]{Julius-Maximilians-Universit{\"a}t W{\"u}rzburg, Fakult{\"a}t für Physik und Astronomie, Institut für Theoretische Physik und Astrophysik, Lehrstuhl für Astronomie, Emil-Fischer-Str. 31, D-97074 W{\"u}rzburg, Germany}

\affiliation[2]{Max-Planck-Institut f\"ur Radioastronomie, Auf dem H\"ugel 69, 53121 Bonn, Germany}

\affiliation[3]{South African Radio Astronomy Observatory, Farm 502 JQ, Broederstroom Road, Hartebeesthoek, 1740, South Africa}

\affiliation[4]{Wits Centre for Astrophysics, School of Physics, University of the Witwatersrand, 1 Jan Smuts Avenue, Johannesburg, 2000, South
Africa}

\affiliation[5]{CSIRO Space and Astronomy, PO Box 76, Epping, NSW 1710, Australia}
\affiliation[6]{NASA HQ, 300 E St SW, DC 20546-0002, Washington DC, USA}

\abstract{The SKA-MPIfR telescope (SKAMPI) is a prototype SKA-Mid antenna located at the SKA site in the Karoo Desert, South Africa. It is funded by the Max Planck Society, through the Max Planck Institute for Radio Astronomy (MPIfR), and operated in collaboration with the South African Radio Astronomy Observatory (SARAO).

The first fringe-finding experiments have been conducted with the European VLBI Network and the southern hemisphere Long Baseline Array, connecting  SKAMPI  with Europe and Australia. Here we present early SKAMPI {\it VLBI mode} results in S-Band, including successful fringe detections and evaluating the integration and imaging performances of SKAMPI in observations with the EVN, LBA, and VLBA. 
}

\begin{document}
\maketitle

\section{Introduction}

The SKA-MPIfR demonstrator dish (SKAMPI) is a prototype radio telescope of the Square Kilometre Array (SKA) Mid frequency  (SKA-Mid) and is operated by the Max Planck Institute for Radio Astronomy (MPIfR) as a stand-alone observatory. SKAMPI serves as a technology demonstrator and pathfinder for SKA-Mid antennas and the MeerKAT plus. It allows some key components and concepts to be validated under realistic conditions. The experience gained helps to reduce technical risks and informs design choices for SKA-Mid. The SKAMPI was fully assembled in mid-2018 at the South African SKA site in the Karoo semi-desert, where the nearest SKA-Mid dishes, SKA060 and SKA107, are situated about 800 to 1500 meters away. Initial test observations took place in December 2019, and technical commissioning, such as system evaluation, radio-frequency-interference testing and performance testing took place until early 2022, leading to the publication of the SKA system design qualification documents in 2022. One of the important results from this early test is that SKAMPI's surface accuracy of 0.3 mm enables observation up to 50 GHz or Q-band \ref{section:EDD}. Since then, developments have continued to establish a framework to operate SKAMPI remotely, to integrate telescope operations with frontend and backend control, and to synchronize observations with data acquisition and automated calibration. 
The current telescope control system is developed on top of the Effelsberg direct digitization (EDD) backend and provides a highly automated system for operations, including observation execution, initial data processing, and transfer of the data products from South Africa to Bonn/Effelsberg for long-term archiving or other post-processing locations.

Furthermore, the EDD system offers different observing modes such as the Very Long Baseline Interferometry (VLBI) mode, 
that offers a unique opportunity for VLBI observations connecting to the available VLBI networks in the world, enabling the study of radio sources at milliarcsecond (mas) resolutions. The feasibility of such observations is investigated and the first technical tests are conducted and will be presented in this chapter.

\section{Systems}

SKAMPI is equipped with an S-Band (1.75 - 3.5~GHz) and a Ku-Band (2~GHz bandwidth within 11 to 18~GHz) linear and circular polarized receiver systems respectively. In combination with the capabilities of the EDD backend it offers different observing modes, such as dual polarisation or full Stokes spectroscopy, a pulsar pipeline mode including pulsar timing, search and baseband recording, and a VLBI mode. As such, it is  an ideal test facility for technical and science programs.

\subsection{The Effelsberg Direct Digitization (EDD) System}
\label{section:EDD}
The EDD system is a flexible and scalable framework for radio astronomy observing systems, covering the full signal chain from the receivers and analogue signal conditioning to time distribution, digitisation, packetisation, and backend signal processing. Developed by the MPIfR, the EDD was initially designed to modernize the signal path of the Effelsberg 100-m telescope and has since evolved into a general-purpose digital backend architecture. It provides direct sampling of intermediate-frequency (IF) signals and real-time data transport using commercial off-the-shelf (COTS) components, such as high-speed network switches, graphics processing units (GPU), and field-programmable gate arrays (FPGA). The modular and reconfigurable backend architecture can be adapted to various observing modes and receiver configurations (e.g., spectroscopy, pulsar timing, beamforming, correlation, VLBI). The software is able to support simultaneous operation of multiple observing modes --- for instance, VLBI recording in parallel with spectrometer output for local monitoring and Radio Frequency Interference (RFI) diagnostics.

The SKAMPI EDD digitises the analogue IF outputs 
at ADC sampling rates of up to 6 Giga Samples per second and resolutions of 8, 10, or 12 bits. A Global Positioning System (GPS)-disciplined hydrogen maser provides the reference clock and 1 pulse per second timing signals to digitizers, ensuring high-precision time stamping. The data are packetised and transmitted as multicast User Datagram Protocol streams across the backend network, allowing multiple subscribing signal-processing pipelines to access them simultaneously.  For VLBI operations, the backend performs digital down-conversion on GPUs to transform the wideband data into multiple VLBI-compliant sub-bands (e.g. $2^N$ MHz bands). These down-converted streams are captured by a dedicated recording pipeline, formatted according to the VLBI Data Interchange Format (VDIF) standard, and written to files on disk together with additional non-VDIF metadata headers. Future developments will include real-time calibration support and direct data streaming to VLBI correlators for e-VLBI applications.

In general, the EDD system design minimizes the need for dedicated hardware and simplifies system maintenance. 

\subsection{The SKA Prototype Dish System / Telescope Control}
\begin{figure}
\includegraphics[width=\textwidth]{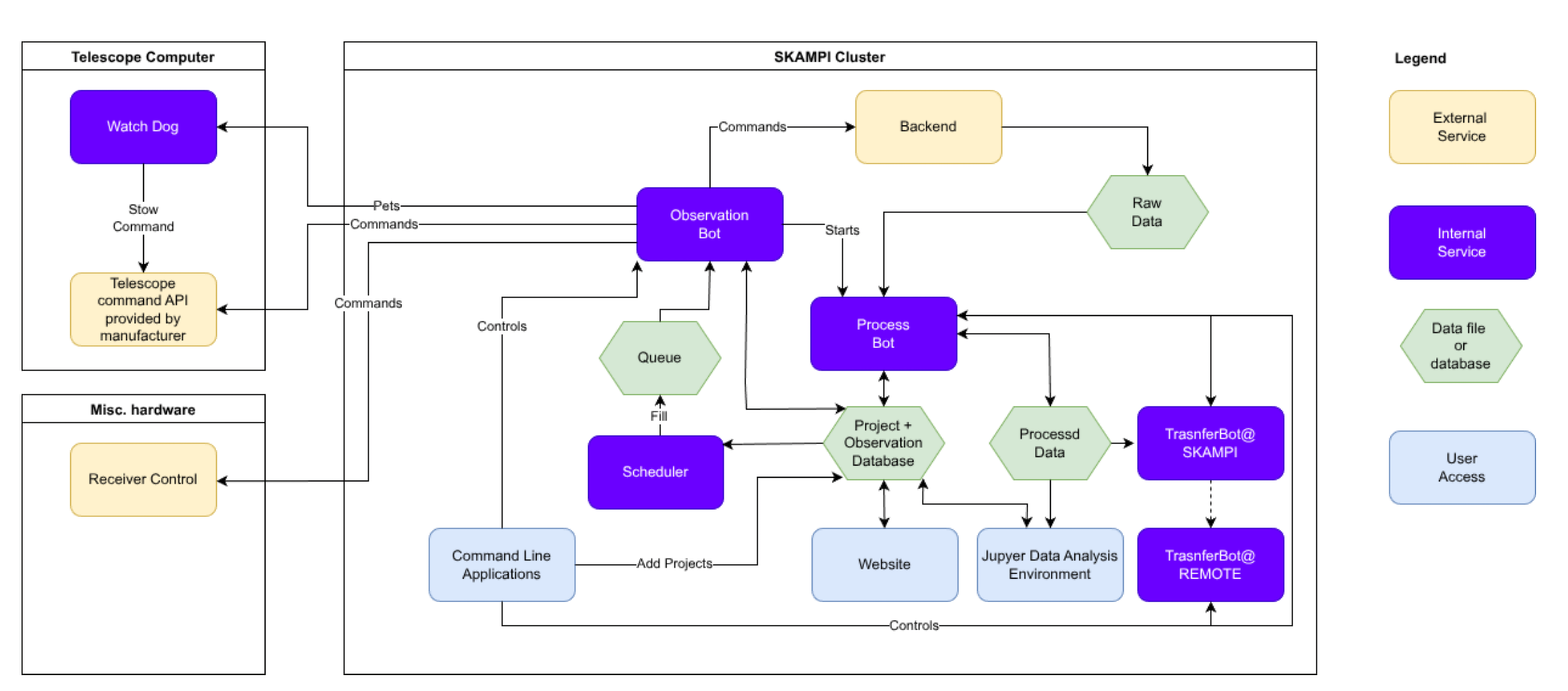}
\caption{\label{fig:skampi_soft} Overview of SKA-MPI Software Architecture: Shown are the individual components of the system, the color coding indicates the type and arrows indicates the communication between the objects.
}
\end{figure}

SKAMPI is operated by a specially developed telescope control and (pre-) processing data acquisition system. The system is intended to enable highly autonomous telescope operations over extended periods of time with little to no human supervision.
As the EDD is deployed as the sole backend, the system is partially integrated into the EDD. However, adaptation 
and extension to other backends would be easily possible. An overview of the software architecture is given in Figure ~\ref{fig:skampi_soft}. The software is structured to be comprised of loosely coupled individual services in a microservice architecture. 

The ObservationBot commands all operations necessary to collect scientific data to the respective devices. It commands the telescope, receiver(s), and the
backend depending on an observation description obtained from a queue. After
the observation, initial processing tasks are placed on the processing queue
specific to the configured observation type. The ObservationBot itself is
controlled via a terminal-user-interface application using the \texttt{katcp}\footnote{https://github.com/ska-sa/katcp-python}
protocol. A webpage displays high-level information of the ObservationBot and telescope status, the observation queue, and also the recorded data and diagnostic plots, generated by the ProcessBot, are also accessible through this webpage.

The ProcessBot executes tasks from a processing queue, such as merging data
from auxiliary sources with the data recorded by the backend, plotting
integrated and dynamic spectra, running RFI mitigation routines, evaluating the
pointing accuracy and calibration, etc. However which task will be executed is specified in the observing queue by the user, respectively also the
observation mode. The ProcessBot is based on the\texttt{ celery}\footnote{https://github.com/celery/celery}  distributed task queue system, ensuring that multiple workers can process tasks in parallel if needed. Interfacing to High-Throughput Computing (HTC) processing systems such as  HTCondor is envisaged in the future for e.g. larger compute infrastructure, facilitating scalable reprocessing of the data in particular during the development phase of new projects. All processing information and metadata of observations are stored in a central Structured Query Language (SQL) database and can be queried to list the observations by type, project, source, quality criteria, and calibration information.

User access to data and on-site computing resources, a Jupyter hub~\cite{}  instance (an open-source multi-user web-based interactive development environment with emphasis on data science) has been deployed. 
The environment provides the means for in-depth data analysis, the documentation of the analysis as well as data exploration via graphical tools.  While access to the setup on-site is limited to selected people only for security reasons, the setup could be used as a template for an off-site setup, providing access to data and computing resources to scientists and students.

SKAMPI has collected so far more than 11000 hours of data in a semi-autonomous mode with unsupervised operations for up to 14 days. The maximum period of
operation was so far limited by dedicated tests or downtime induced by other
works on the prototypes.

SKAMPI designed storage capacity is upto 180 TB, with write capacity of more than 12 GB/s, enough to process baseband data. The SKAMPI accepts proposals to obtain joint VLBI observation on a best effort basis. Although the logistics to the correlator needs to be organised by the proposer.

\subsection{SKAMPI in various VLBI arrays}
\label{sec:res}
\begin{figure}[h]
    \centering
        \includegraphics[width=0.75\columnwidth]{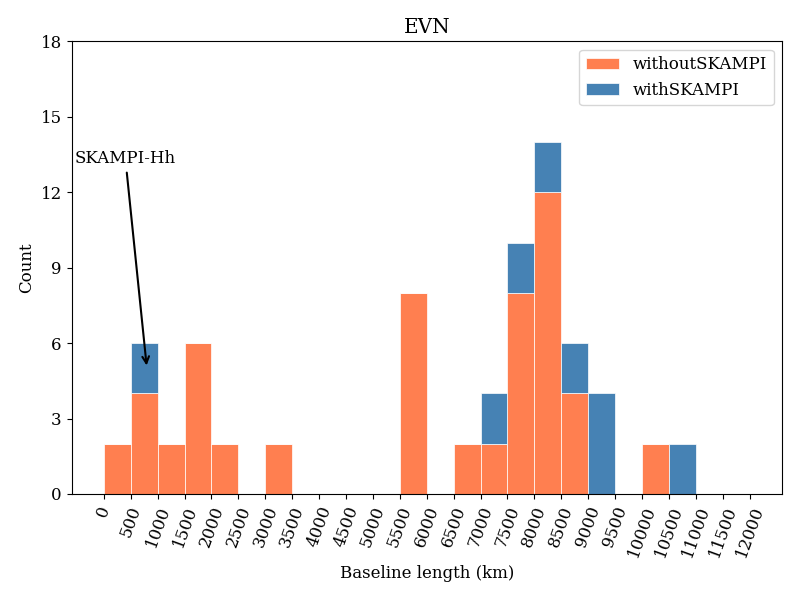}
    \caption{Distribution of baseline lengths for EVN with and without SKAMPI. This histogram shows the frequency of different baseline lengths (in km) formed between SKAMPI and other EVN stations, highlighting the range of angular resolutions achieved. The SKAMPI-HartRAO baseline (814.0 km) is similar to the Ef-Mc baseline (757.0 km), providing redundancy for amplitude calibration. There are a total of 45 unique non-zero baselines among the 10 stations included in the network.}
    \label{fig:EVNhist}
\end{figure}

\begin{figure}[h!]
    \centering
        \includegraphics[width=0.75\columnwidth]{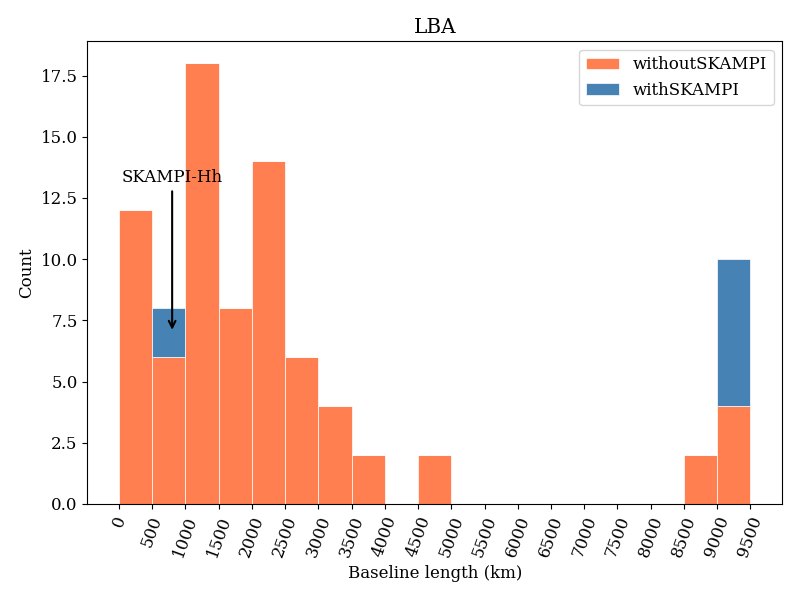}
    \caption{Distribution of baseline lengths for SKAMPI with the LBA. This histogram illustrates the frequency of different baseline lengths (in km) formed between SKAMPI and other LBA stations, showcasing the extent of angular resolutions achieved. The SKAMPI-Hh baseline (814.0 km) is similar to the Td-Ho baseline (831.8 km), providing important redundancy. There are a total of 55 unique non-zero baselines among the 11 stations included in the network.Note that, SKAMPI doubles the number of baselines, which are critical for high angular resolution.}
    \label{fig:LBAhist}
\end{figure}

 In this section, we will demonstrate SKAMPI’s role in various VLBI arrays. We will focus on the case of 2~GHz (S-band) observations to address the current possible resolution, assuming 24-hour VLBI sessions for sources in a favorable declination ($\delta$) of $\delta$ = $-10$ deg, See Figure \ref{fig:uvEVN}. The simulation has been performed using \texttt{EHTim} \citep{2019ascl.soft04004C}. The position and System Equivalence Flux Density (SEFD) of each station is obtained from the EVN Observation Planner\footnote{https://planobs.jive.eu}. For SKAMPI, the SEFD at S-band was assumed to be 360 Jy, based on the first measurement (Jünemann et al., in prep.). Baselines used in the simulation are shown in Figure \ref{fig:EVNhist},\jw{and} \ref{fig:LBAhist} to demonstrate the visibilities for both the unresolved point source and the double source.

\subsubsection{SKAMPI with the European VLBI network (EVN) and Long Baseline Array (LBA)}

\begin{figure}[h!]
    \centering
	\includegraphics[width=1.0\columnwidth]{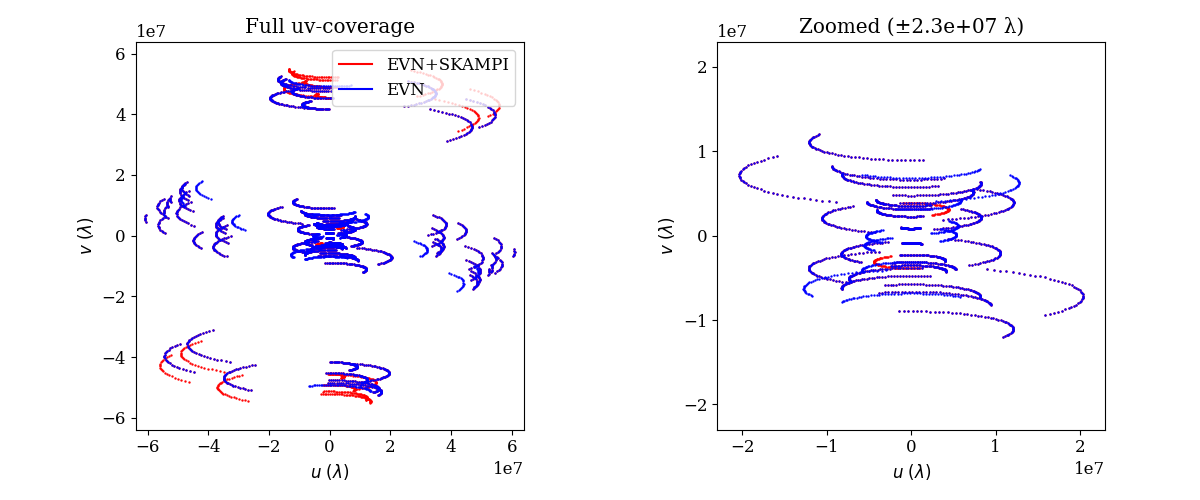}
    \caption{The ($u,v$) coverage from simulated 24 hours observation at 2~GHz for an EVN observation with and without SKAMPI.  \jw{The simulated source is located at $\delta$ = -10$^\circ$.} Left: SKAMPI will double the number of rare data points in the outermost part of the ($u,v$) domain and even provide the longest baselines (red). Right: SKAMPI-Hart (red) baselines will even the same parts of the $uv$ plane in a similar manner as inner-European baseline, which adds to the calibration stability of the array.}
    \label{fig:uvEVN}
\end{figure}

\begin{figure}[h!]
    \centering
	\includegraphics[width=1.0\columnwidth]{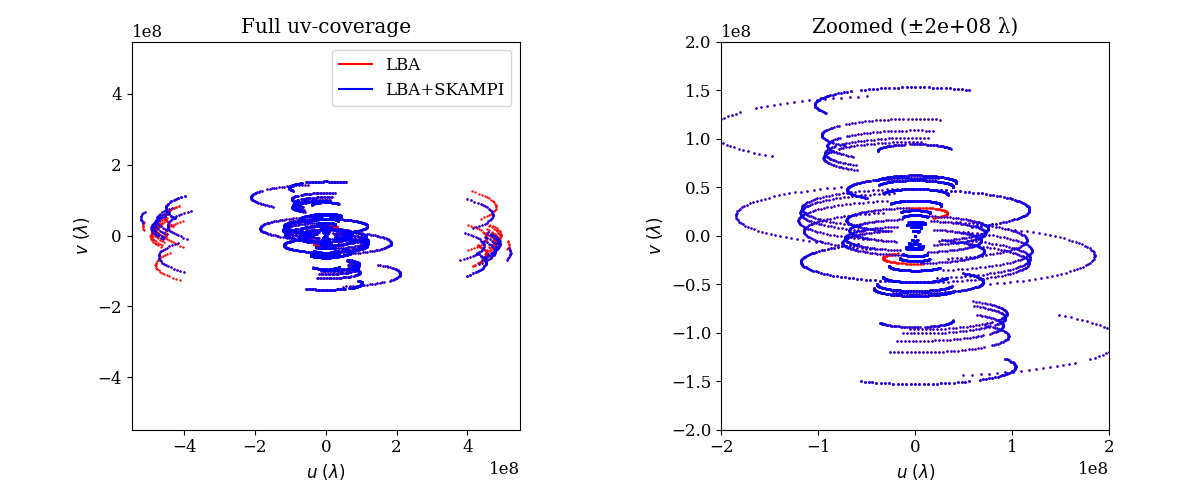}
    \caption{The ($u,v$) coverage from simulated 24~hours observation at 2~GHz for an LBA observation with and without SKAMPI. \jw{The simulated source is located at $\delta$ = 10$^{\circ}$.} Left: Showing that SKAMPI will 
    double the number of rare data points in the outermost part of the ($u,v$) domain and provide the second-longest baselines. Right: SKAMPI-Hart baselines willwill even the same parts of the $uv$ plane in a similar manner as inner-European baseline, which adds to the calibration stability of the array.}
    \label{fig:uvLBA}
\end{figure}

SKAMPI provides an additional African VLBI station. It increases flexibility on the longest east–west baselines of the LBA and on the north–south baselines of the EVN, which in the past has relied heavily on Hartebeesthoek (HartRAO, Hh) as the sole African element. 
The coverage of the critical parts of the longest-baseline section of the ($u,v$) plane has therefore been extremely sparse, limiting the image fidelity.
Furthermore, the amplitude calibration on the longest baselines was solely dependent on the treatment of baselines involving data from the Hh station.
This can be overcome by the addition of a new (South-)African antennas to VLBI arrays.
The SKAMPI–HartRAO baseline ($\sim$814 km) is comparable to Effelsberg–Medicina ($\sim$760~km) and Tidbinbilla–Hobart ($\sim$830~km), so SKAMPI not only strengthens the array geometry but also adds redundancy for amplitude cross-checks with HartRAO (see red tracks  in Figures \ref{fig:uvEVN} and \ref{fig:uvLBA}).

\begin{figure}[h!]
    \centering
	\includegraphics[width=0.45\columnwidth]{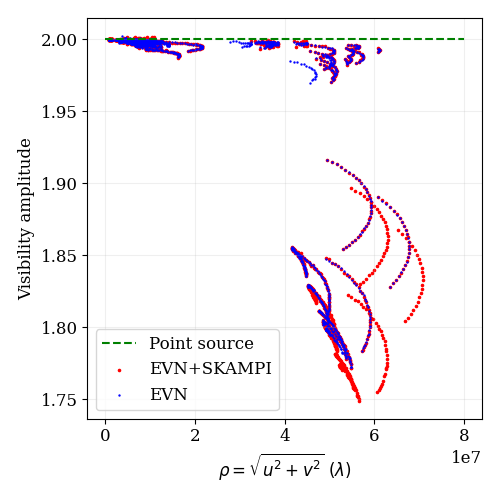}
    \includegraphics[width=0.45\columnwidth]{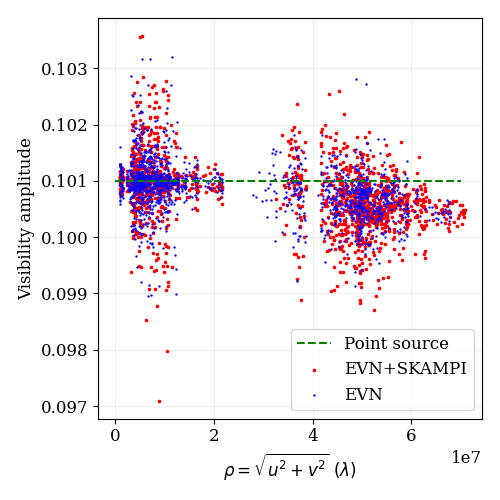}
    \caption{(Left) Simulated visibility amplitude from EVN and EVN with SKAMPI. The object used in this simulation is a double point source with 1~Jy flux density at 0.1~mas separation in the N-S direction from each other. This demonstrates that the source can be distinguished from a point source even at an angular separation corresponding to 20\% of the nominal resolution. (0.4 mas) with more than 20\% amplitude difference to the point source case (constant line at 2~Jy, green dash line). (Right) Similar setup but with the source flux density of 1~mJy and 0.1 Jy, where now the visibility amplitude appeared almost indistinguishable (deviation less than 1\%) from a point source scenario.}
    \label{fig:simEVN}
\end{figure}

To demonstrate this we simulate two point sources of 1~Jy each with a separation of 0.1~mas. We found that SKAMPI with  a few EVN stations, which are Effelsberg (Ef), Yebes (Ys), Medicina (Mc), Urumqi (Ur), Noto (Nt), Westerbork (Wb), Kunming (Km), and Tianma (Tm) and Hartebeesthoek (Hh), can already confirm the nature of the source being an extended source even with a really small separation, as the amplitude-$uv$ distance represents \jw{absolute value} the Fourier transform of the image. If the source is a point source, the plot would be just a constant line. However, if the source is a double source, Figure \ref{fig:simEVN} shows both resolved and unresolved visibilities showing significantly different visibility patterns.

\section{Fringe Tests}

\subsection{EVN and SKAMPI} 
The first successful SKAMPI fringe test was carried out on 2024~Feb~19, observing blazars 3C454.3 and 0420$-$014 at 2~GHz together with Effelsberg, Medicina, and Yebes in a 2.5~hours long VLBI experiment (project code: {\em te019}). Backend systems were the EDD at SKAMPI, Digital Base Band Converter (DBBC) at the other stations. Two 2~MHz wide channels were recorded in dual polarization.

Baseband recordings were transferred to the Bonn MPIfR VLBI computing cluster and were correlated using the DiFX software correlator \citep{2011PASP..123..275D}. Baseband data from the station backends are in a DiFX-compatible VDIF format. However, one conversion step is necessary for SKAMPI data, as they are encapsulated in an EDD-specific format, Distributed Acquisition and Data Analysis\footnote{https://psrdada.sourceforge.net/} (DADA) files, that contain the actual VDIF. Custom headers in the DADA files needed to be stripped to reattain VDIF. Lastly, coordinates used for SKAMPI were the a-priori World Geodetic System 1984 (WGS84) position of the axis intersection of (X,Y,Z) = (5110074.5366~m, 2003957.8440~m, -3239607.6895~m) of unknown precision, with an assumed axis offset of zero. While neither accounts for the parabolic offsetted design of the telescope, they were sufficiently accurate to produce fringes with low residual fringe rates over the course of the observation.

\begin{figure}[p!]
    \centering
	\includegraphics[width=1.0\columnwidth,trim={0cm 20.5cm 0cm 2cm},clip]{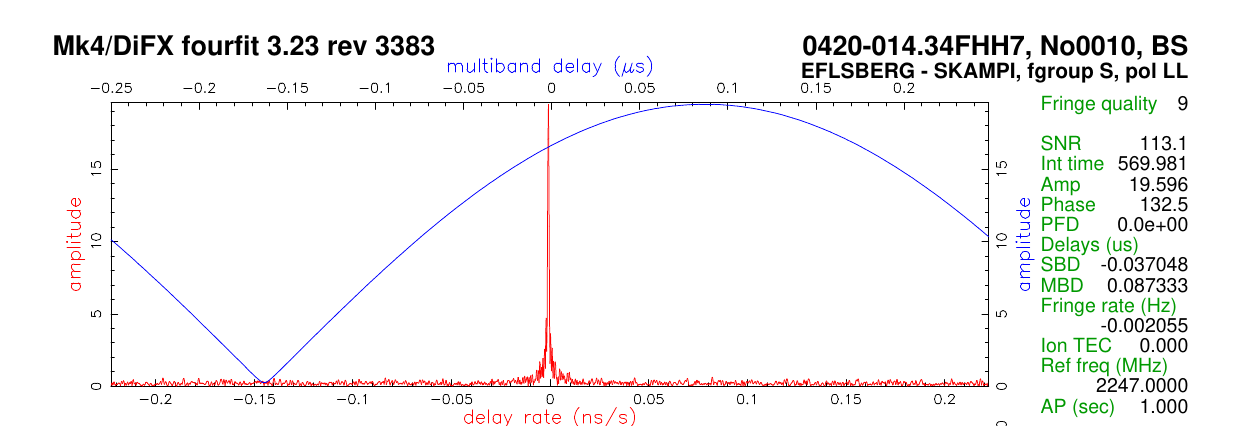} \\
	\includegraphics[width=1.0\columnwidth,trim={0cm 20.5cm 0cm 2cm},clip]{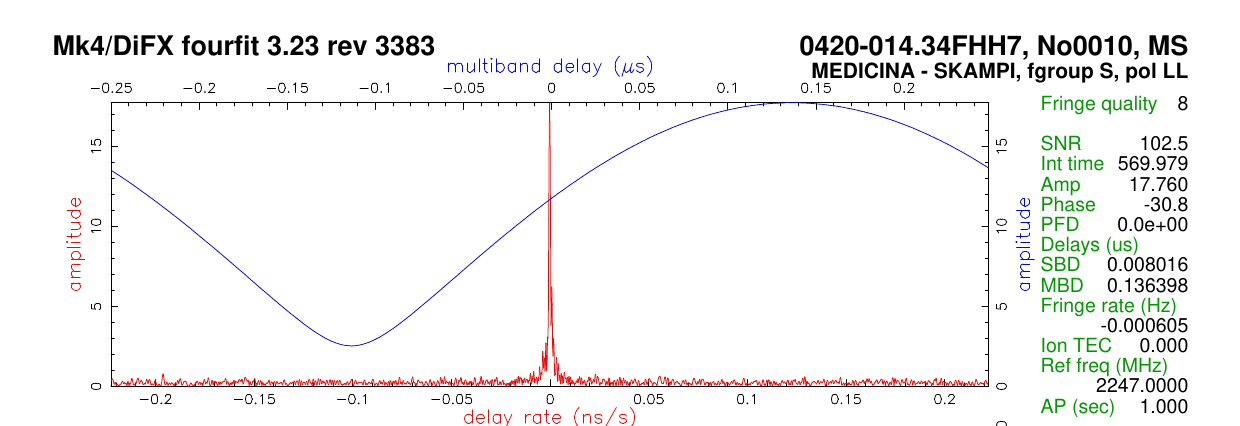} \\
	\includegraphics[width=1.0\columnwidth,trim={0cm 16cm 0cm 2cm},clip]{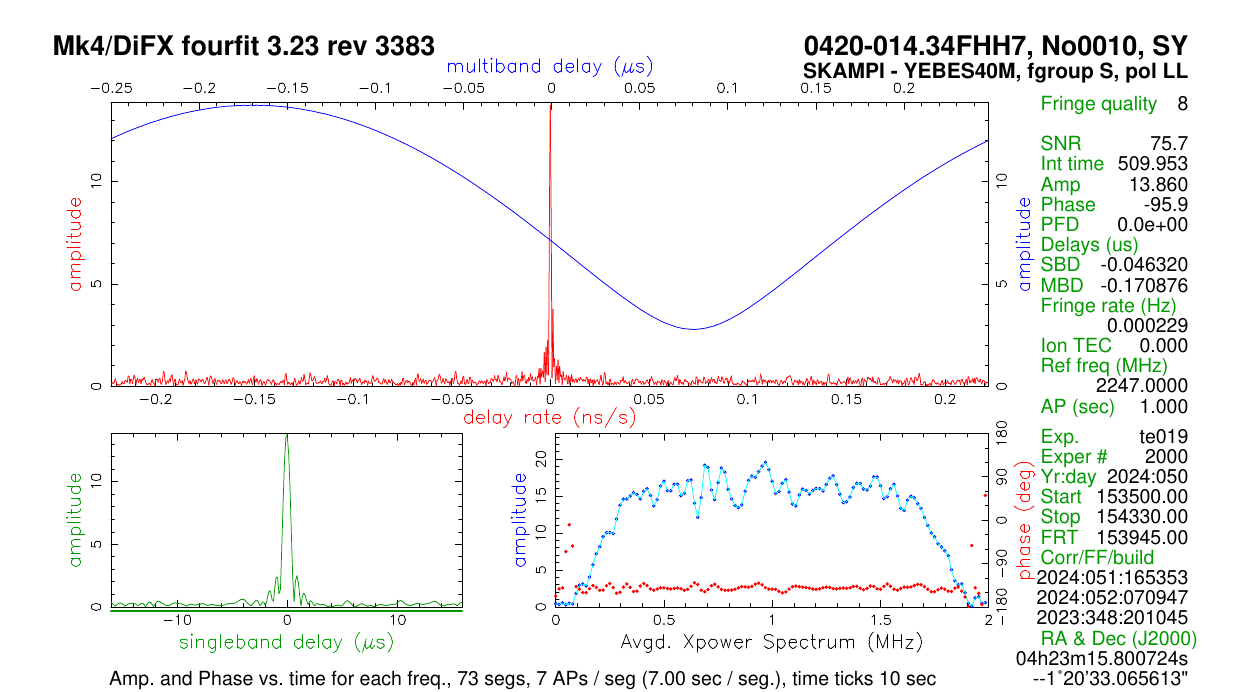}
    \caption{Fringes on blazar 0420-014 in the 2~GHz SKAMPI-EVN VLBI test observation of 2024 February 19.  Residual delay rate spectra (red) and multi-band delays (blue) of the fringes on three EVN baselines to SKAMPI are shown in the top panels. 
    The lower panels show the residual delay of the SKAMPI-Yebes fringe (bottom left, green), and the averaged bandpass amplitude (bottom right, light blue) and phase (red) on that baseline. The high fringe SNRs of ca. 75 to 110 and the narrow peaks in delay and delay rate spectra indicate good sensitivity and good atmospheric and instrumental phase stability over time.}
    \label{fig:first_fringes_fig}
\end{figure}

The first SKAMPI fringes at 2~GHz are shown in Figure~\ref{fig:first_fringes_fig}. The a-priori coordinates for SKAMPI were accurate for the fringe finding. They can be refined in dedicated future geodetic-like observations. The attained high fringe signal-to-noise ratios (SNRs) and clean fringe spectra (cf. strong narrow peaks without spread in Fig.~\ref{fig:first_fringes_fig}) \jw{hints at} good atmospheric observing conditions and good SKAMPI VLBI system performance. Future technical tests can characterize the phase stability of the receiver and the efficiency of the EDD VLBI backend in more detail.

\subsection{SKAMPI with LBA} 
Following the EVN fringe test, the first SKAMPI-LBA experiment on 2024 May 24 at S-band was performed as a part of the regular TANAMI \citep[Tracking Active Galactic Nuclei with Austral Milliarcsecond Interferometry,][]{2010A&A...519A..45O,2024A&A...681A..69B}. The aim was to verify end-to-end interoperability of scheduling, recording, transfer, correlation, and mixed-polarization handling between SKAMPI and the LBA.

\begin{figure}[htb!]
    \centering
	\includegraphics[width=1.0\columnwidth,trim={1cm 2cm 6cm 0cm},clip]{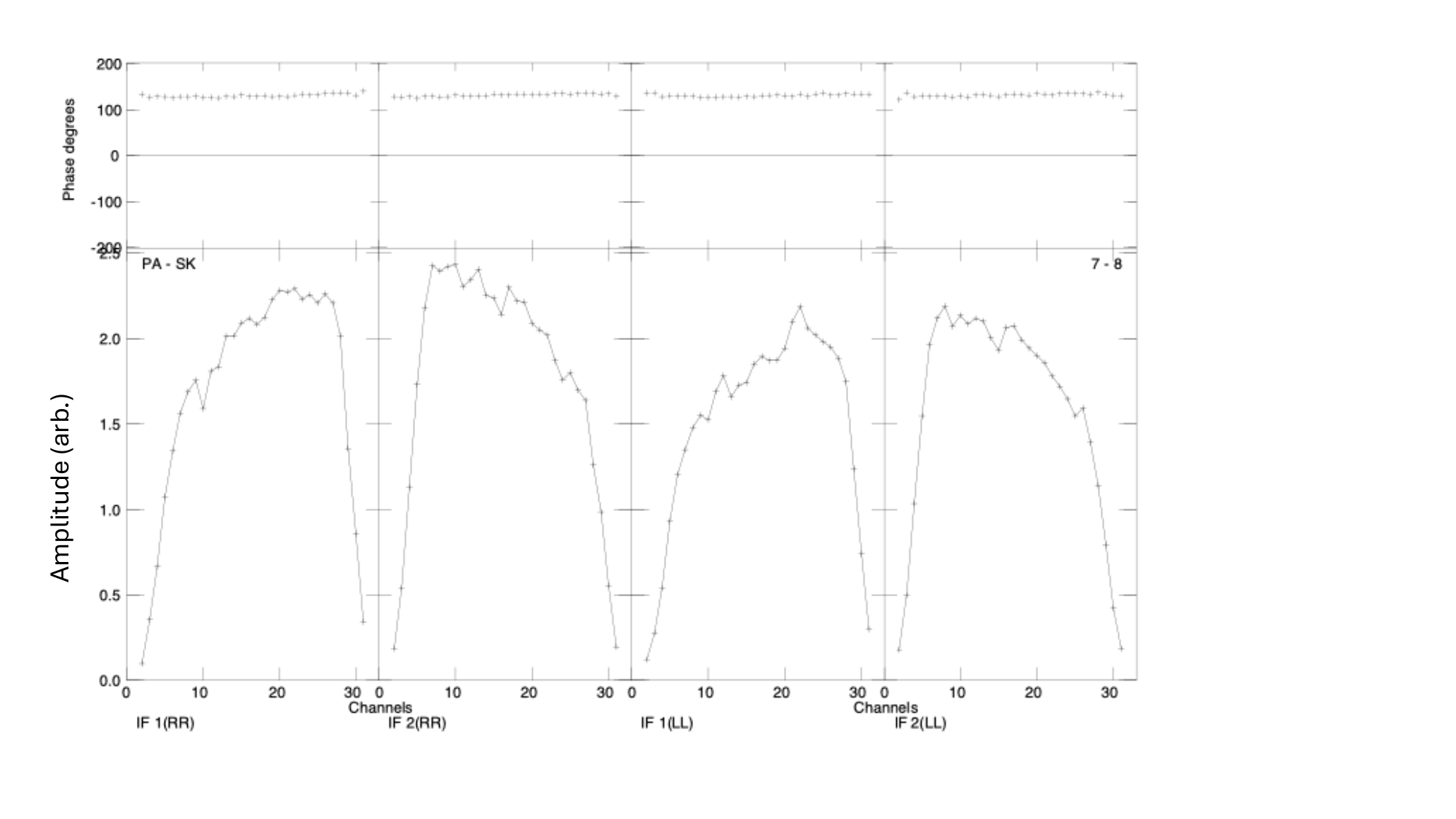} \\
	
    \caption{LBA observations baseline (SKAMPI $-$ Parkes) phase and amplitude versus channels plot (upper and lower) for IF1 (RCP), IF2 (RCP), IF1 (LCP), and IF2 (LCP) from left to right, respectively, at 2~GHz on the 2024 May 24 TANAMI observation. The results indicate successful fringe finding: for both IFs, the phase varies linearly across the bands, while the amplitude shows a smooth profile across the channel.}
    \label{fig:LBA_fringes_fig}
\end{figure}

With VDIF data from EDD, the additional header were removed to match the VDIF format.The data were transferred with \texttt{jive5ab}\footnote{https://github.com/jive-vlbi/jive5ab} from SKAMPI to a staging server at HartRAO and then forwarded to the CSIRO correlator at the Pawsey Centre for Supercomputing Research in Australia.\\
Correlation was carried out with the DiFX software correlator, and SKAMPI’s linear-feed data were converted to circular after correlation using \texttt{PolConvert} \citep{imvidal2016}. During the S-band fringe search, WGS84 site coordinates and a zero axis offset were good enough approximations to detect fringes. \\
We detected fringes after searching over a wide delay and time window, and the test verified end-to-end operation and compatibility between SKAMPI and the LBA at S-band. Figure~\ref{fig:LBA_fringes_fig} shows  phase and amplitude against frequency (as 1024 channels over 2 MHz) of the SKAMPI-Parkes cross power spectrum. The results indicate successful fringe finding: for both IFs, the phase varies linearly across the bands, while the amplitude shows a smoothprofile across the channel. The need for a wide search was traced to an uncorrected station clock offset in the EDD system, which also made later the step of the calibration harder. 

\section{Future outlook and Conclusions}

\begin{figure}[h!]
    \centering
	\includegraphics[width=1.0\columnwidth]{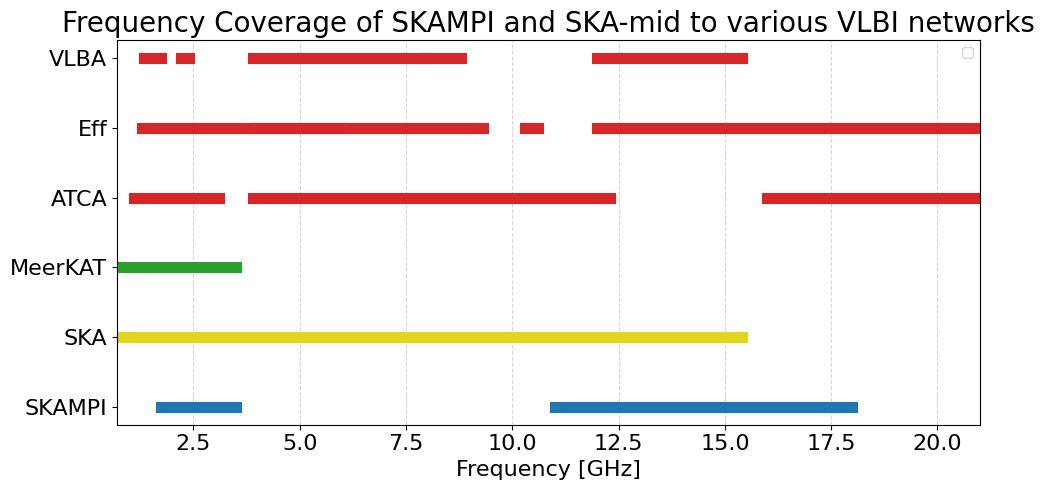} \\
	
    \caption{Frequency coverage of SKAMPI to SKA and Effelsberg (Eff; represents the EVN) and ATCA (represents the LBA frequency capability) showing missing frequency coverages between SKAMPI and these instruments. The frequency coverage of the SKA-Mid precursor, MeerKAT, also shown. 
    }
    \label{fig:SKAMPI_FREQ_COVERAGE}
\end{figure}

\subsection{VLBI Use-Cases for SKAMPI}
\label{sec:skampi_vlbi_usecases}

Due to its unique location, SKAMPI can act as an individual station offset from the SKA-Mid core array or can be a substitute station  for VLBI observations that require SKA baselines but not the sensitivity of the full array to achieve the same science goal \citep[see][for details]{2019arXiv190110361P}. We outline examples VLBI science cases enabled by SKAMPI operating as an individual station in VLBI networks in this subsection. 

For the frequency coverage, we analyze the current and future VLBI network frequency coverages and the SKA-Mid \citep{braun2019anticipatedperformancesquarekilometre}\footnote{https://www.skao.int/en/science-users/118/ska-telescope-specifications}. Currently, SKAMPI will only overlap with SKA-Mid Band2 over a narrow band of 10 MHz (1.75-1.76 GHz), with SKA-Mid Band 5b over a 3.55 GHz band (11.85-15.4 GHz) and with band3 over 1.3 GHz (1.8 - 3.1 GHz). To compare that with various VLBI networks, we obtain the frequency coverage of VLBA from the frequency and band performance page\footnote{https://science.nrao.edu/facilities/vlba/docs/manuals/oss/bands-perf} while we use the Australia Telescope Compact Array (ATCA)\footnote{www.narrabri.atnf.csiro.au/observing/users\_guide/html/atug.html} and Effelsberg\footnote{eff100mwiki.mpifr-bonn.mpg.de/doku.php?id=information\_for\_astronomers:rx\_list} frequency coverage to represent LBA and EVN capabilities. Figure~\ref{fig:SKAMPI_FREQ_COVERAGE} displays the frequency coverages of the SKA-Mid, SKAMPI, and VLBI networks.

\paragraph{Southern sky VLBI.} First, as shown in Section \ref{sec:res}, SKAMPI adds long baselines that will improve angular resolution and closure quantities for compact, high-brightness sources and hotspots. This supports structure characterization and model-fitting for compact radio sources by imaging at S and Ku bands to study the compactness, opacity, and astrometry. Moreover, making use of both S and Ku receivers, the phenomenon where the core of the AGN jet is shifted with the observing frequency (“core-shift”) and core variability can be studied by joining SKAMPI to existing AGN programs such as TANAMI.

\paragraph{Rapid-response ToO VLBI with an Africa--Europe mini-array.}
Similar to EVN-lite \citep{2022evlb.confE..35M}, SKAMPI can operate as an element of \jw{a} Target-of-Opportunity array with partners in southern Africa, including SKA-Mid, Hartebeesthoek, and potentially Ghana telescope \citep[see][]{Bempong-Manful01.2026.SKA}. Fast triggering and low-latency correlation deliver first-epoch localisations and structural constraints for the southern source. This can be further improved by enhancing this array with some stations of the LBA and/or EVN.

\subsection{Conclusions}

The first VLBI measurement from the SKA-Mid site, using the SKAMPI VLBI observing mode was tested in 2024 using telescopes in Spain (Yebes), Germany (Effelsberg), and Italy (Medicina). The data from these telescopes have been correlated at the Bonn DiFX VLBI correlator and fringe plots of amplitude and phase correlations as functions of delay, delay rate, and time have shown a stable connection to all observatories and fringes even at the  longest baselines of 8800~km (SKAMPI–Effelsberg). In addition, SKAMPI participated in test observations with the Southern hemisphere Long Baseline Array as part of the TANAMI program. The fringe finding was successful and shows that SKAMPI can be connected to such long baselines. In order to fully integrate the VLBI capabilities into the SKAMPI observing mode, further developments are being pursued,  such as automated scheduling of observations, scientific calibration, and integration of SKAMPI parameters into the common software suites of the VLBI community\footnote{See gitlab.mpcdf.mpg.de/twinchen/ska\_mpg\_prototype\_processing/-/tree/master/OBS\_MODE\_VLBI}. We anticipate opening
SKAMPI for proposals to the South African and German communities starting at
the end of 2026.

In the near future, the construction of the Botswana telescope (Boss) at the Botswana International University of Science and Technology will provide a short baseline between Hh and Boss (approximately 300~km), while simultaneously offering similar baselines between SKAMPI-Hh and SKAMPI-Boss. This enhanced setup will offer more tools for more flexibility in calibrations schemes, thereby adding more stability to the flux calibration process.

{\bf Acknowledgement}:\\ SKAMPI, the SKA-MPG prototype telescope, is a facility of the Max-Planck Society (MPG) and was established with the assistance of the South African Radio Observatory (SARAO). It is jointly operated and maintained by the Max Planck Institute for Radio Astronomy (MPIfR) and SARAO. This research was made possible with the support of the MPIfR and SARAO. This article is based on observations carried out with the Yebes 40 m telescope. The 40 m radio telescope at Yebes Observatory is operated by the Spanish Geographic Institute (IGN; Ministerio de Transportes y Movilidad Sostenible). The Medicina radio telescope is funded by the Ministry of University and Research (MUR) and is operated as National Facility by the National Institute for Astrophysics (INAF).  We acknowledge financial support by the German Federal Ministry of Education and Research (BMBF) under ErUM-Pro grant 05A23WW3 (Verbundprojekt D-MeerKAT III).

The Long Baseline Array is part of the Australia Telescope National Facility\footnote{https://ror.org/05qajvd42}, which is funded by the Australian Government for operation as a National Facility managed by CSIRO. This work was supported by resources provided by the Pawsey Supercomputing Research Centre with funding from the Australian Government and the Government of Western Australia

\bibliographystyle{AASKAII_draft_template_v1/abbrvnat-maxbibnames4}
\bibliography{AASKAII_draft_template_v1/chapter}

\end{document}